\documentclass[aps,prb,floatfix,twocolumn,amsmath,amssymb]{revtex4}
\usepackage{graphicx}
\usepackage{dcolumn}
\usepackage{bm}
\usepackage{color}
\usepackage{soul}
\usepackage{nicefrac}

\begin{document}

\title{Non-local parity order in the two-dimensional Mott insulator}
\author{Serena Fazzini,$^{1}$ Federico Becca,$^{2}$ and Arianna Montorsi$^{1}$}
\affiliation{
 $^{1}$ Institute for condensed matter physics and 
 complex systems, DISAT, Politecnico di Torino, 
 I-10129, Italy \\
 $^{2}$ CNR-IOM-Democritos National Simulation Centre
 and International School for Advanced Studies (SISSA),
 Via Bonomea 265, I-34136, Trieste, Italy}

\date{\today}

\begin{abstract}
The Mott insulator is characterized by having small deviations around the (integer) average particle density $n$, with pairs 
with $n-1$ and $n+1$ particles forming bound states. In one dimension, the effect is captured by a non-zero value of a non-local 
``string'' of parities, which instead vanishes in the superfluid phase where density fluctuations are large. Here, we investigate 
the interaction induced transition from the superfluid to the Mott insulator, in the paradigmatic Bose Hubbard model at $n=1$. 
By means of quantum Monte Carlo simulations and finite size scaling analysis on $L \times M$ ladders, we explore the behavior 
of ``brane'' parity operators from one dimension (i.e., $M=1$ and $L \to \infty$) to two dimensions (i.e., $M \to \infty$ and 
$L \to \infty$). We confirm the conjecture that, adopting a standard definition, their average value decays to zero in two 
dimensions also in the insulating phase, evaluating the scaling factor of the ``perimeter law'' [S.P. Rath {\it et al.}, 
Ann. Phys. (N.Y.) {\bf 334}, 256 (2013)]. Upon introducing a further phase in the brane parity, we show that its expectation 
value becomes non-zero in the insulator, while still vanishing at the transition to the superfluid phase. These quantities are 
directly accessible to experimental measures, thus providing an insightful signature of the Mott insulator.
\end{abstract}

\maketitle

{\it Introduction.} The theoretical prediction of exotic orders in quantum phases of matter~\cite{Haldane1983,Affleck1987,Nijs1989} 
has been followed, in recent years, by the attempt of their realization in quantum gases of ultracold atoms.~\cite{Endres2011} 
A deep understanding of the role played by {\it non-local orders} (NLOs) and their intertwined connection with long-range 
entanglement and topological features represent the next fundamental questions.~\cite{Wenbook} The Mott insulator (MI), 
which is induced by interaction in both fermionic and bosonic models, is a paradigmatic example of a quantum phase that has no 
classical counterparts. Indeed, unlike various ordered phases (e.g., displaying charge or spin order), it is not described by 
the presence of any long-range order, identified with a non-zero value in the asymptotic limit of the two-point correlation 
function of an appropriate local observable. Nevertheless, in the one-dimensional bosonic Hubbard model, the MI was characterized by 
a non-vanishing NLO, defined in terms of a {\it parity operator} that acts along a one-dimensional string of sites.~\cite{Berg2008} 
The existence of such order was measured in the insulating phase of a gas of ultracold bosonic atoms, confined into lattices 
with a strong one-dimensional anisotropy.~\cite{Endres2011} Then, it was realized that NLO also characterizes the MI of fermionic 
Hubbard models.~\cite{Montorsi2012,Barbiero2013} Indeed, the general picture of the MI consists of a state with an integer number 
of particles per site $n$ ($n=1$ for the fermionic case) in which the relevant excitations are given by holons (sites with $n-1$ 
particles) and doublons (sites with $n+1$ particles).~\cite{NoteHD} These quantum fluctuations form bound pairs with a finite 
correlation length in the MI, so that their presence does not change the overall parity of a string of sites unless for the pairs 
separated by its boundary, which represent a zero-measure set. The gapless phase, which for bosons is also superfluid (SF), is 
reached when the correlation length of the pairs becomes infinite. Such behavior is expected to take place in arbitrary 
dimensions.~\cite{Capello2007,Capello2008} In Ref.~\onlinecite{Rath2013}, it was argued that in spatial dimensions greater than 
one, the qualitative change occurring at SF-MI transition could be captured by an appropriate two-dimensional generalization 
(the ``brane'') of the one-dimensional string of parities. However, in the thermodynamic limit this quantity is vanishing in both 
the SF and MI phases (with different asymptotic behaviors), making elusive its experimental detection. More recently it was 
conjectured that a properly normalized brane parity could remain finite in the MI phase, also in higher dimensional (fermionic) 
systems.~\cite{Boschi2016}

In this letter, by computing brane parity operators within a numerically exact quantum Monte Carlo technique, we study both the 
SF and MI in the Bose Hubbard model. Our results confirm that the mechanism underlying the SF-MI transition does not change
going from one to two dimensions. Moreover, since the standard brane parity NLO vanishes in two dimensions, we generalize it by 
introducing an arbitrary phase $\theta$.~\cite{Qin2003} For an appropriate choice of $\theta$, the NLO is proved to remain finite 
also in the two-dimensional MI, while being still vanishing in the SF phase. Finally, by comparing the numerical results with 
analytic approximations of the parity, we characterize the difference between the MI and the SF phase in terms of the fundamentally 
different behavior of the density fluctuations.

{\it The model.} The Bose Hubbard model~\cite{BHM} on ladders with $L \times M$ sites reads:
\begin{equation}\label{eq:bosehubbard}
{\cal H} = -\frac{t}{2} \sum_{\langle R,R^\prime \rangle} b^\dag_{R} b^{\phantom{\dagger}}_{R^\prime} 
+ {\rm h.c.} + \frac{U}{2} \sum_R n_R (n_R-n),
\end{equation}
where $\langle R,R^\prime \rangle$ indicates nearest-neighbor sites, $b^\dag_R$ ($b^{\phantom{\dagger}}_R$) creates (destroys) a 
boson on the site $R$, and $n_R=b^\dag_R b^{\phantom{\dagger}}_R$ is the density on the site $R$. The density per site is fixed to 
be $n=N_b/N_s$, where $N_b$ and $N_s=L \times M$ are the number of bosons and sites, respectively. In the following, we concentrate
on the case with $n=1$. We indicate the coordinates of the sites with $R=(x,y)$ and consider periodic-boundary conditions in both 
directions (except for the case with $M=1$ and $2$, for which open-boundary conditions are considered along the rungs). In order to 
assess the properties of the two-dimensional limit, we first fix $M$ and perform the extrapolations for $L \to \infty$ and then 
increase $M$. Thus, by varying the number of legs $M$ (and extrapolating $M \to \infty$), we are able to get insights into the 
two-dimensional case. 
 
{\it Brane parities with phases.} In general, we can define the {\it on-site} parity operator ${\cal P}_R$ that describes the 
density fluctuations at the site $R$ with respect to its average value $n$, namely ${\cal P}_R= {\rm e}^{i \pi (n_R-n)}$. Depending 
on the parity of the boson density $n_R$ with respect to $n$, we have that ${\cal P}_R= \pm 1$. In one-dimension (i.e. $M=1$), 
the non-local parity is defined as the string of on-site parity operators (from $x=0$ to $x=r$):
\begin{equation}\label{eq:NLO1D}
O_P(r,M=1)=\prod_{0 \le x < r} {\cal P}_{x,0}.
\end{equation}
This definition can be extended to the case with $M>1$ in various ways. In particular, we can introduce a {\it brane} of on-site 
parity operators:
\begin{equation}\label{eq:NLO2D}
O_P(r,M)=\prod_{0 \le x < r} \prod_{0 \le y < M} {\cal P}_{x,y} =
\prod_{0 \le x < r} {\cal P}^{\rm rung}_{x}(M),
\end{equation}
where ${\cal P}_x^{\rm rung}$ is defined in terms of the {\it rung density} $n^{\rm rung}_x = \sum_{y=0}^{M-1} n_{x,y}$, i.e., 
${\cal P}^{\rm rung}_x(M)= {\rm e}^{i \pi (n^{\rm rung}_x-Mn)}$. 

\begin{figure}
\includegraphics[width=\columnwidth]{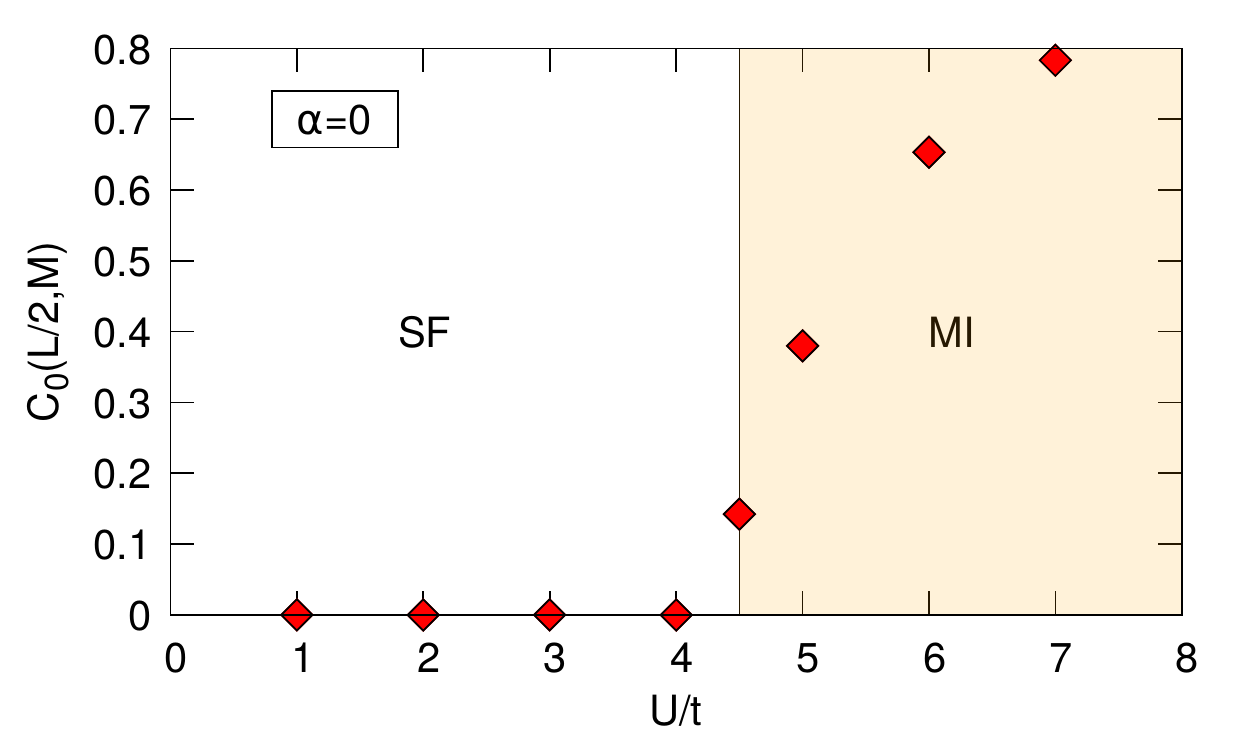}
\caption{\label{fig:CPM2}
(Color online) Brane parity correlator $C_0(r,M)$, evaluated at $r=L/2$, as a function of $U/t$ for ladders with $M=2$ and $L=120$.}
\end{figure}

\begin{figure}
\includegraphics[width=\columnwidth]{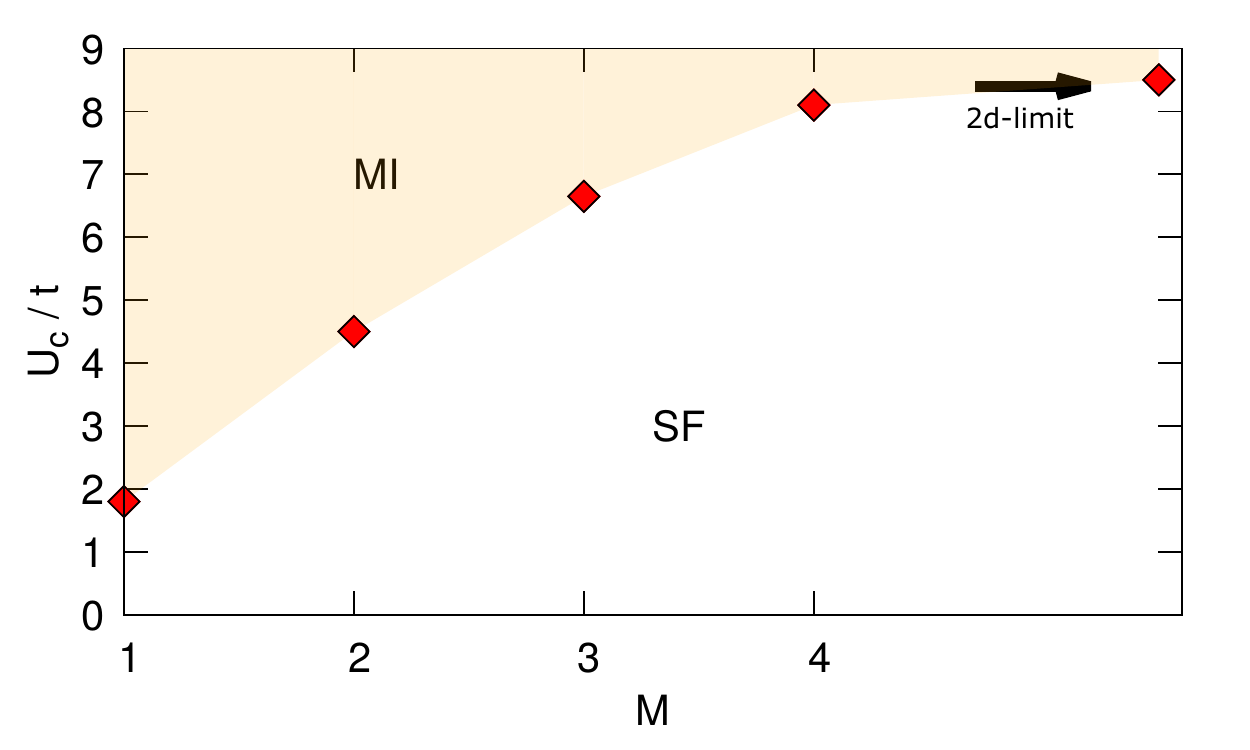}
\caption{\label{fig:phasediag}
(Color online) Phase diagram of the bosonic Hubbard model for $n=1$: the critical interaction strength $U_c$ at which the 
superfluid-Mott transition occurs is reported as a function of the number of legs $M$ of the ladder.}
\end{figure}

In one dimension ($M=1$), SF and MI phases can be distinguished by looking at the ground-state expectation value of $O_P(r,M)$:
\begin{equation}
C_P(r,M) \equiv \langle O_P(r,M) \rangle = \langle\Psi_0| O_P(r,M) |\Psi_0\rangle,
\end{equation}
which coincides with the correlation function $\langle O_P^\dagger(0,M) O_P(r,M) \rangle$. Indeed, $C_P(r,1)$ is known to give a 
finite value for $r \to \infty$ in the MI, while it is vanishing in the SF phase of bosons~\cite{Berg2008} or in the metallic phase 
of fermions,~\cite{Montorsi2012} thus playing the role of an order parameter for the MI phase. For higher spatial dimensions, the 
situation is more subtle. In fact, for bosons in two dimensions, it has been argued~\cite{Rath2013} that $C_P(r,M)$ should decay to 
zero with $M$ and $r \to \infty$ in both the MI and SF phases; however, a different asymptotic behavior should appear in these two 
cases (see below). Recently a generalization of the brane parity operator~(\ref{eq:NLO2D}) has been suggested, which has a non-vanishing 
expectation value in the MI also in the $M \to \infty$ limit, thanks to a normalization with the number of legs $M$ of the phase 
in $P_x^{\rm rung}(M)$.~\cite{Boschi2016} In this regard, we observe that the density fluctuations on a rung of length $M$ can be 
associated to a ``spin'' of length $2M+1$ (both for fermions and for bosons, in the latter case when only small fluctuations with 
$n-1$ and $n+1$ particles are considered). Then, the Hamiltonian on the $M$-legs ladder can be associated to a spin-$M$ model on 
a chain. In analogy with the choice made in the latter case for the Haldane string operator,~\cite{Oshikawa1992,Qin2003} we can 
generalize the brane parity operator by introducing an arbitrary phase $\theta$, and define:
\begin{equation}\label{eq:OPth}
O_P^{(\theta)}(r,M) \equiv \left [ O_P(r,M) \right ]^\frac{\theta}{\pi},
\end{equation}
where $\theta$ depends on $M$ and possibly the model Hamiltonian. In particular, in case of the Heisenberg model, one obtains that 
$\theta=\frac{\pi}{M}$ maximizes the average value of the parity string operator, which is also the result found in 
Ref.~\onlinecite{Boschi2016} for the MI on a fermionic ladder.

More generally, we suggest that, for appropriate values of $\theta$, the expectation value of the generalized parity operator:
\begin{equation}\label{eq:par}
C_P^{(\theta)}(M) = \lim_{r \to \infty} \langle O_P^{(\theta)} (r,M)\rangle
\end{equation}
could behave as order parameter for the SF-MI transition of the Hubbard model also in two dimensions (i.e., for $M \to \infty$), 
remaining asymptotically finite in the MI, while vanishing in the SF phase. In order to test this conjecture, we give a first 
derivation of the behavior of $C_P^{(\theta)}(M)$ for the bosonic Hubbard model within a Gaussian approximation. In this case, 
we obtain:
\begin{equation}\label{NLP2D}
\langle O_P^{(\theta)} (r,M) \rangle \approx e^{-\frac{\theta^2}{2}\langle\delta N^2\rangle},
\end{equation}
where $\delta N=\sum_{x=0}^{r-1}(n_x^{\rm rung}-Mn)$ describes the total density fluctuations on the brane of size $r$. 
The latter can be evaluated generalizing Ref.~\onlinecite{Rath2013}:
\begin{equation}
\label{eq:Gauss}
C_P^{(\theta)}(M) \approx
\begin{cases} 
\lim_{r\to \infty} r^{-a M \theta^2} & {\rm SF}, \\
e^{-b M \theta^2}                    & {\rm MI},
\end{cases}
\end{equation}
where $a$ and $b$ are (positive) constants related to the physical parameters. Thus, assuming $\theta \propto M^{-\alpha}$, we
have that $C_P^{(\theta)} = \lim_{M \to \infty} C_P^{(\theta)}(M)$ is finite within the MI for $\alpha \ge \frac{1}{2}$. By contrast, 
for $\theta=\pi$ (i.e., $\alpha=0$), we recover the ``perimeter-law'' decay found in Ref.~\onlinecite{Rath2013} (here, $2M$ is 
the perimeter of the brane enclosed in $O_P(r,M)$). Within the SF phase $C_P^{(\theta)}(M)$ is zero at any finite $M$ for arbitrary 
$\theta$. Noticeably, the value $C_P^{(\theta)}=0$ is independent on the order of the two limits $M \to \infty$ and $r \to \infty$
(i.e., $L \to \infty$) only for $\alpha \leq \frac{1}{2}$.

{\it Numerical results.} In the following, by considering extensive Monte Carlo simulations on ladders with different values of $M$, we 
will study the actual behavior of the generalized parity operator~(\ref{eq:par}) for $\theta=\pi/M^{\alpha}$ and $\alpha=0$, $1/2$, and
$1$ in the bosonic Hubbard model~(\ref{eq:bosehubbard}). In order to simplify the notation, for finite values of $r$ and $M$, we define
\begin{equation} 
C_{\alpha}(r,M) \equiv \langle O_P^{(\frac{\pi}{M^\alpha})} (r,M)\rangle,
\end{equation} 
$C_{\alpha}(M)$ its limiting value for $r \to \infty$, and $C_\alpha \equiv \lim_{M\rightarrow\infty} C_\alpha(M)$.
The ground-state properties of the Hamiltonian are obtained by using the Green's function Monte Carlo technique.~\cite{Trivedi1990} 
In particular, we used the algorithm with fixed number of walkers; moreover, observables, such as generalized brane parities, are 
computed by using the so-called forward-walking technique.~\cite{Calandra1998}

\begin{figure}
\includegraphics[width=\columnwidth]{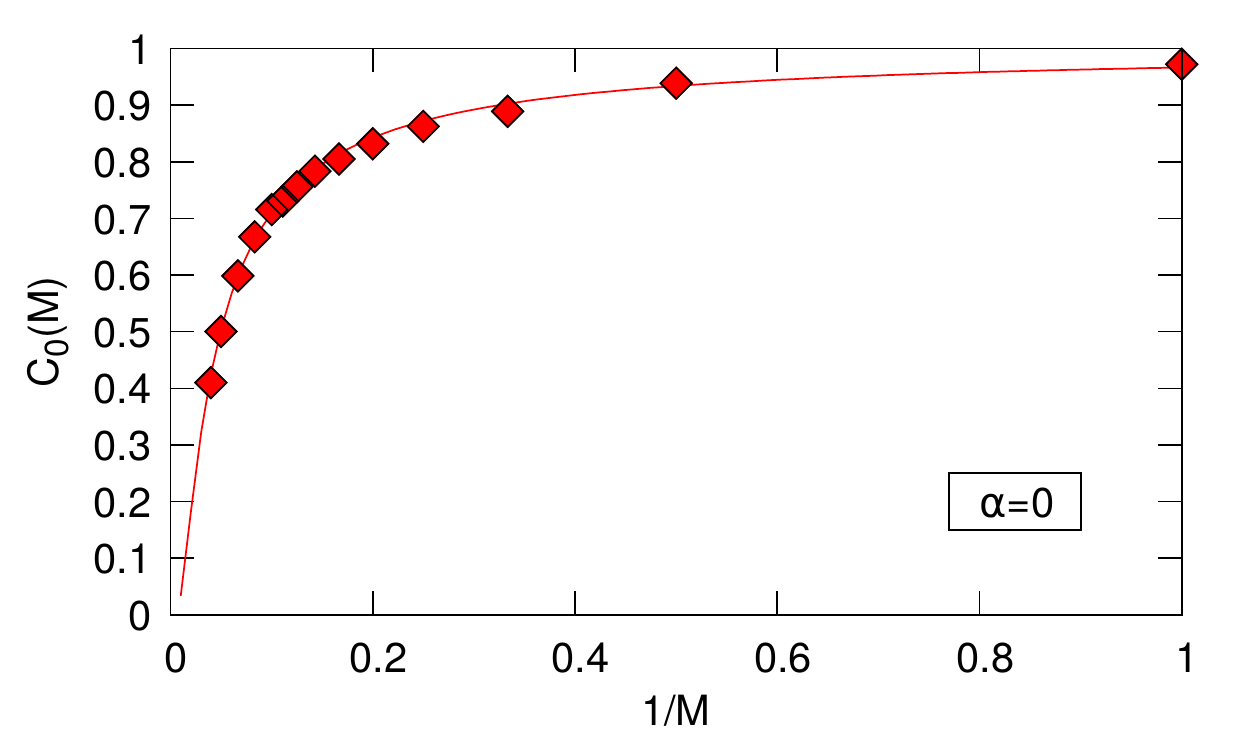}
\caption{\label{fig:scalingIP}
(Color online) Size scaling of brane parity $C_0(M)$ (i.e., $\theta=\pi$) with $M$ for $U/t=12$. The fit is performed by using 
Eq.~(\ref{eq:Gauss}) with $b=t^2/(2U^2)$. The results have been obtained for ladders with $L=30$, after having verified that the 
calculations do not change sensibly for larger values of $L$.}
\end{figure}

\begin{figure}
\includegraphics[width=\columnwidth]{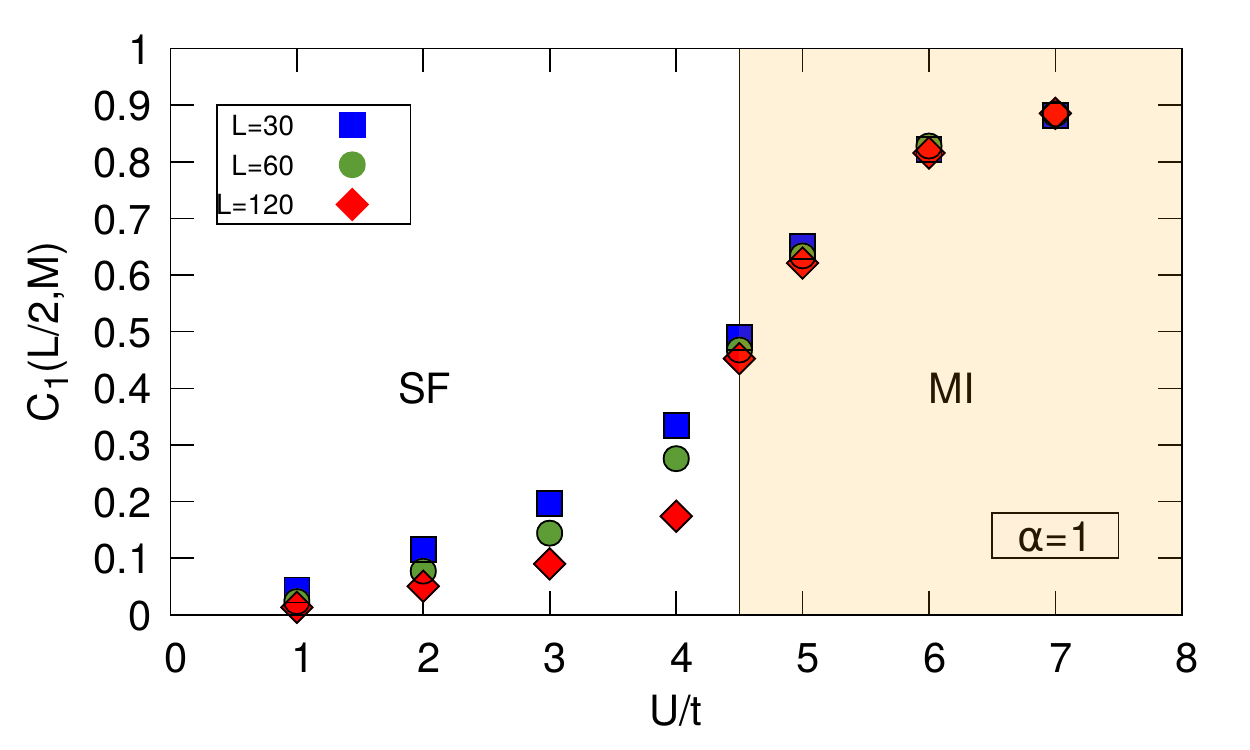}
\caption{\label{fig:CFM2}
(Color online) Brane parity correlator $C_{1}(r,M)$ (i.e. $\theta=\pi/M$), evaluated at $r=L/2$, as a function of $U/t$ for ladders 
with $M=2$ and various lengths $L$.}
\end{figure}

First of all, we study the SF-MI transition by computing the parity operators for several values of $M$. In this way, we obtain a 
rather precise determination of the critical value of $U_c$ when increasing the number of legs, from the one-dimensional case up to 
the two-dimensional limit. For a ladder with fixed $L$ and $M$, we evaluate $C_{\alpha}(r,M)$ at $r=L/2$. We first consider the case of 
the standard parity, i.e. $\alpha=0$. In Fig.~\ref{fig:CPM2}, we show its behavior as a function of $U/t$ for a ladder with $M=2$ and 
$L=120$. Here, $C_0(L/2,M=2)$ is vanishing for small values of the interaction strength and becomes finite when increasing $U/t$, 
signaling the transition between the SF and the MI phases. We notice that the transition point signaled in the figure has been located 
after having performed the asymptotic limit $L \to \infty$, i.e., after having computed $C_0(M=2)$. Based on the results on $C_0(M)$, 
once the thermodynamic limit $L \to \infty$ (for each value of $M$) has been performed, we can draw a phase diagram in which we report 
the critical point $U_c$ for different values of $M$, see Fig.~\ref{fig:phasediag}. We would like to emphasize that the transition point 
is monotonically increasing with $M$ and converges quite rapidly to the value obtained in two dimensions.~\cite{Capello2007,Capogrosso2008}
Indeed, we find that $U_c/t=1.8(1)$ in one dimension, while it is already $U_c/t=8.1(1)$ for $M=4$, to be compared with the value of 
$U_c/t=8.5(1)$ that has been obtained in two dimensions.

\begin{figure}
\includegraphics[width=\columnwidth]{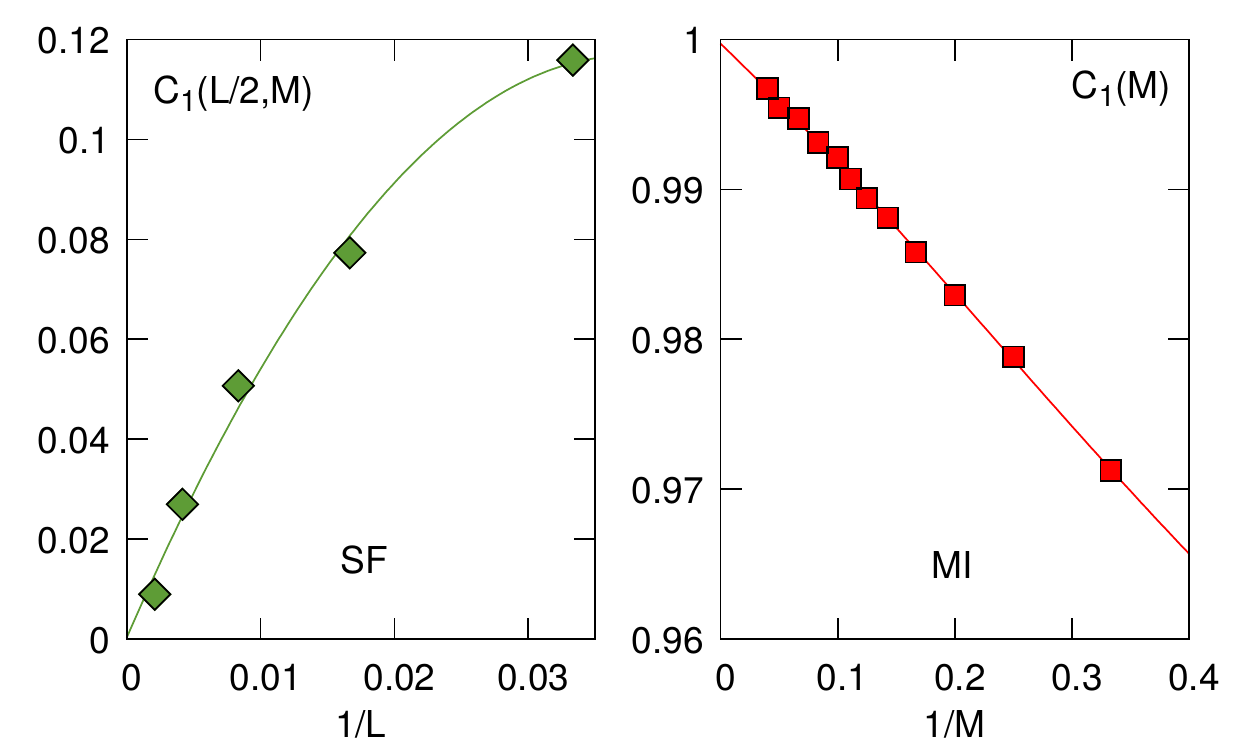}
\caption{\label{fig:scalingFP}
(Color online) Left panel: Finite-size scaling of $C_1(L/2,M)$ for $M=2$ with increasing $L$, deep inside the SF regime (i.e., $U/t=2$).
Right panel: Finite size-scaling of {$C_1(M)$} with increasing the number of legs $M$ of the ladder, deep inside the MI phase (i.e., 
$U/t=12$). Here, the results have been obtained for ladders with $L=30$, after having verified that the calculations do not change 
sensibly for larger values of $L$.}
\end{figure}

\begin{figure}
\includegraphics[width=\columnwidth]{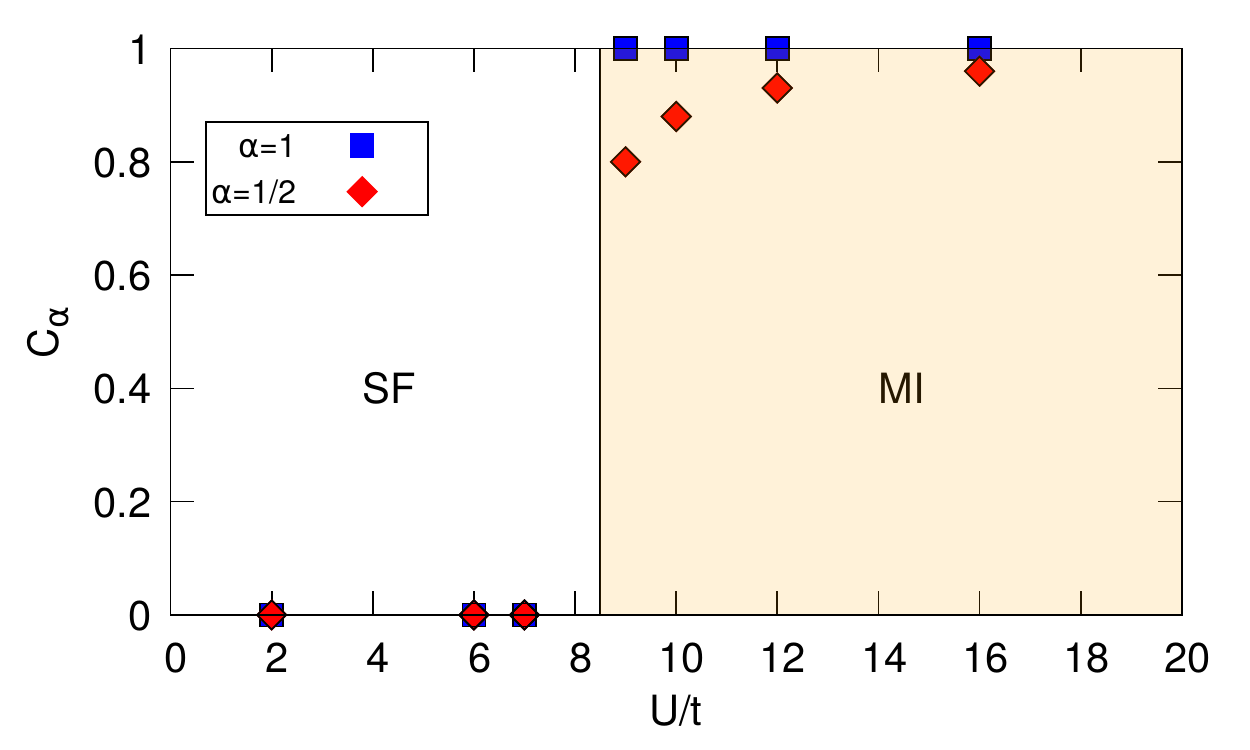}
\caption{\label{Fig:Comparison}
(Color online) Two-dimensional brane parity $C_{\alpha}$ as a function of $U/t$ for $\alpha=1$ (i.e., $\theta=\pi/M$) and $\alpha=1/2$ 
(i.e., $\theta=\pi/\sqrt{M}$).}
\end{figure}

Even though $C_0(M)$ is finite in the MI for any {\it finite} value of $M$, its value decreases to zero when $M \to \infty$, in agreement 
with what has been predicted in Ref.~\onlinecite{Rath2013}. For example, in Fig.~\ref{fig:scalingIP} we report the size scaling of $C_0(M)$ 
for $U/t=12$ deep inside the MI. There the results have been obtained for ladders with $L=30$, after having verified that the calculations 
do not change sensibly for large values of $L$. In particular, we find that our data can be fitted by using Eq.~(\ref{eq:Gauss})
with $b=t^2/(2U^2)$. In this respect, a totally different scenario appears when considering the brane parity with $\alpha=1$. We still 
obtain that for any finite values of $M$ $C_1(M)$ vanishes within the SF regime, while it is finite within the MI,(see Fig.~\ref{fig:CFM2}
and left panel of Fig.~\ref{fig:scalingFP}). Most importantly, $C_1(M)$ remains finite within the MI also when increasing the number of 
legs $M$ to the two-dimensional limit, as shown in the right panel of Fig.~\ref{fig:scalingFP} where $C_1$ is extrapolated. In fact, we 
numerically verified that $C_1=1$ for each value of $U$ in the MI phase, as suggested again by the Gaussian approximation of 
Eq.~(\ref{eq:Gauss}). The latter also predicts that, in the case $\alpha=1/2$, $C_{1/2}$ is finite in the two-dimensional MI 
phase, in this case with a non-trivial dependence on $U$, e.g., $C_{1/2}=e^{-\pi^2 b}$. Our numerical simulations confirm this behavior, 
as shown in Fig.~\ref{Fig:Comparison} in which we compare the two cases with $\alpha=1$ and $\alpha=1/2$ in the two-dimensional limit. 
We would like to mention that, within the SF phase, the modified parity operator $C_1(r=L/2,M)$ shows very large size effects 
when extrapolating $L \to \infty$ (at fixed $M$); these size effects are indeed, much larger than those observed for the standard parity 
(e.g., compare Figs.~\ref{fig:CFM2} and~\ref{fig:CPM2}). This fact is again in agreement with Eq.~(\ref{eq:Gauss}). 

{\it Conclusions.} We have addressed the issue of characterizing the MI and the transition to the SF phase in more than one-dimensional 
systems. In particular, we explored the capability of generalized brane parity operators to capture the order present in the MI phase. 
By performing Monte Carlo simulations on the bosonic Hubbard model on rectangular clusters with $L$ rungs and $M$ legs, we have 
investigated the asymptotic limit $L \to \infty$, when passing from one dimension ($M=1$) to the two-dimensional case ($M \to \infty$). 
We have shown that the average value of the standard brane parity operator $C_0(M)$ works as an order parameter for the MI at any finite 
$M$. However, it decays to zero with a ``perimeter law'' when considering two spatial dimensions~\cite{Rath2013}, thus rendering elusive 
its experimental measure. By contrast, exploiting the fact that in the MI small fluctuations around the average density $n$ take place, 
and pairs with $n+1$ and $n-1$ are strongly correlated, we have argued that a generalized brane parity operator $C_{\alpha}$ is non-zero 
in two dimensions for any $\alpha \geq 1/2$. In fact, in the Mott phase, the Gaussian approximation predicts that $C_\alpha=1$ for 
$\alpha > 1/2$; by contrast, $C_{1/2} \approx {\rm e}^{-\frac{t^2}{2U^2}}$, thus enlightening the role of interaction in driving the 
Mott transition. Moreover, $C_{1/2}=0$ is obtained within the SF. These facts suggest that the proper order parameter to describe the 
MI-SF transition could be $C_{1/2}$. Indeed, our numerical results show a very good agreement with the predictions.~\cite{note2} 
Presently in-situ density fluctuations can be measured in cold atom experiments by means of high-resolution imaging.~\cite{Endres2011}
Our results provide a unique tool to probe in these systems the presence of a MI, and its interaction induced evolution up to the 
transition to the SF phase.

Finally, we would like to make a remark on fermionic Hubbard-like models. In one dimension, the MI phase takes place in the charge 
degrees of freedom, whereas a corresponding Luther-Emery phase, with possibly dominant superconducting correlation, may take place in the 
spin channel. It has already been noticed~\cite{Montorsi2012} that such phase is captured by a parity NLO in which density fluctuations 
are replaced by magnetization ones. We expect that generalized brane parity operators in both density and spin channels, together with 
their Haldane counterparts,~\cite{Oshikawa1992,Barbiero2013} will help to clarify the phase diagram in two-dimensional models, where a 
number of different phases (including magnetic states, spin liquids, and superconductors) can be stabilized by changing the band 
structure and the interaction terms.


\begin{thebibliography}{}
\bibitem{NoteHD} The terminology with ``holons'' and ``doublons'' originates from the fermionic Hubbard model, where excitations 
   have empty and doubly occupied sites.
\end{thebibliography}
\end{document}